\documentclass[fleqn,10pt]{wlscirep}
\usepackage[utf8]{inputenc}
\usepackage[T1]{fontenc}
\usepackage{makecell}
 \usepackage{url}

\pdfoutput=1

\title{Representing Polymers as Periodic Graphs with Learned Descriptors for Accurate Polymer Property Predictions}

\author[1*]{Evan R. Antoniuk}
\author[2]{Peggy Li}
\author[3]{Bhavya Kailkhura}
\author[1**]{Anna M. Hiszpanski}
\affil[1]{Materials Science Division, Physical and Life Sciences Directorate, Lawrence Livermore National Laboratory, Livermore, CA, USA}
\affil[2]{Global Security Computing Applications Division, Computing Directorate, Lawrence Livermore National Laboratory, Livermore, CA, USA}
\affil[3]{Machine Intelligence Group/Center for Applied Scientific Computing, Computing Directorate, Lawrence Livermore National Laboratory, Livermore, CA, USA}

\affil[*]{antoniuk1@llnl.gov}
\affil[**]{hiszpanski2@llnl.gov}

\begin{abstract}
One of the grand challenges of utilizing machine learning for the discovery of innovative new polymers lies in the difficulty of
accurately representing the complex, multi-scale structures of polymeric materials. Although a wide array of hand-designed polymer representations have been explored, there has yet to be an ideal solution for the problems of how to capture the periodicity of polymer structures, as well as how to automatically develop polymer descriptors without the need for human feature design. 
In this work, we tackle these problems through the development of our periodic \textit{polymer graph} representation. Specifically, our pipeline for polymer property predictions is comprised of our \textit{polymer graph} representation that naturally accounts for the periodicity of polymers, followed by a message-passing neural network (MPNN) that leverages the power of graph deep learning to automatically learn chemically-relevant polymer descriptors. Across a diverse dataset of 10 polymer properties, we find that this \textit{polymer graph} representation consistently outperforms previous hand-designed representations with a sizable 20\% average reduction in prediction error. Our results illustrate how the incorporation of chemical intuition through directly encoding periodicity into our \textit{polymer graph} representation is responsible for a considerable improvement in the accuracy and reliability of polymer property predictions. We also demonstrate how combining \textit{polymer graph} representations with message-passing neural network architectures can automatically extract meaningful polymer features that are consistent with human intuition, while achieving better predictive performance than human-derived features. Whereas previous polymer informatics have typically utilized hand-designed descriptors that have been optimized for representing molecules, this work highlights the advancement in predictive capability that is possible if using chemical descriptors that are specifically optimized for capturing the unique chemical structure of polymers.

\end{abstract}
\begin{document}

\flushbottom
\maketitle

\thispagestyle{empty}

\section*{Introduction}
The complex chemical structure of polymers provides chemists with an unprecedented level of chemical design opportunities. As a result, efforts focused on the rational design of polymers has led to their use in a remarkably wide range of applications.\cite{hou_applications_2019,briscoe_chapter_2008} By varying a polymer's monomer structure, its propagation site, or the polymer chain length, the properties of the polymer can be tuned to suit the application of interest.\cite{sharma_rational_2014,pacult_effect_2018,mun_conjugated_2019,selivanova_branched_2019} Formulations of polymers with chemical additives adds an even richer potential design and functional space. However, efficiently navigating this vast chemical design space to identify promising polymer materials for a given application is incredibly challenging. Recently, the field of polymer informatics has shown much promise in tackling this challenge by utilizing machine learning (ML) models to learn the complex structure-property relationships of polymers.\cite{goswami_deep_2021,tao_benchmarking_2021} Through the development of increasingly accurate ML models, one can envision the future possibility of fully computational polymer design workflows whereby polymers can be rapidly designed in silico to satisfy multiple property criteria.
\\

Accurate machine learning models require that the input to the model -- be it an image, text, or chemical structure -- is transformed into a machine-readable format that captures and represents all relevant features of the input. The development of suitable polymer representations to use as inputs for machine learning algorithms has been recognized as one of the grand challenges in the field of polymer informatics.\cite{chen_polymer_2021,patra_data-driven_2022} Unlike molecules that consist of a fixed number of atoms and typically only contain dozens of atoms, a given polymer can consist of a periodic chain of between 10$^0$-10$^6$ repeating chemical sub-units.\cite{yamauchi_two-step_2021} This presents a considerable challenge for creating machine-readable polymer representations since they must encode the periodic nature of the polymer structure and also be compact enough to represent up to millions of atoms. Despite these challenges, previous work on small molecules and crystalline solids has demonstrated that developing physically-inspired representations of atomistic systems is a fruitful direction towards realizing significant improvements in property prediction accuracy.\cite{xie_crystal_2018,chemprop_yang_analyzing_2019,schindler_discovery_2020,nguyen_predicting_2021}
\\

In this work, we develop a new polymer representation that is the first to explicitly capture the periodic nature of polymeric materials. First, we show that the widely used monomer-based representations cannot sufficiently describe the periodic, extended structure of polymers and can result in unreliable property predictions. To solve this issue, we have developed a novel polymer representation that we refer to as a \textit{polymer graph} and that naturally incorporates polymers' periodicity in its construction. Across a diverse spectrum of 10 unique polymer properties, we find that our \textit{polymer graph} representation consistently outperforms previous monomer-based representations with a 20\% average reduction in prediction error. Finally, we demonstrate how our \textit{polymer graph} representations can be combined with a Message Passing Neural Network (MPNN) framework to automatically extract chemically-relevant polymer descriptors directly from a dataset of polymer properties. We find that these MPNN-learned polymer features yield more accurate property predictions than predetermined features that have been hand-crafted by chemists over years. 

\section*{Results} \label{Results}

\subsection*{Introduction to Polymer Graphs}

Recently, a wide range of machine learning (ML) algorithms have been developed to rapidly and accurately predict polymer properties in an \textit{a priori} manner.  \cite{huan_polymer_2016,park_graph_prediction_2022} These previous approaches typically represent polymers by their monomer repeat unit which is then converted into either i) molecular descriptors that includes the presence of molecular fragments and/or property descriptors (such as the molecular weight of the molecule)\cite{kuenneth_polymer_2021,wu_machine-learning-assisted_2019,barnett_designing_nodate} or ii) a molecular graph, where atoms are represented as nodes and chemical bonds are represented as edges (Figure 1a)). \cite{park_graph_prediction_2022} It should be noted that these representations were first developed for molecules before being transferred to the field of polymer informatics. Although these monomer-based representations perform well at representing finite molecules, we show in this work that monomer-based representations cannot adequately represent the extended chain structures that define polymers.
\\

\begin{figure}
\centering
\includegraphics[width=\textwidth]{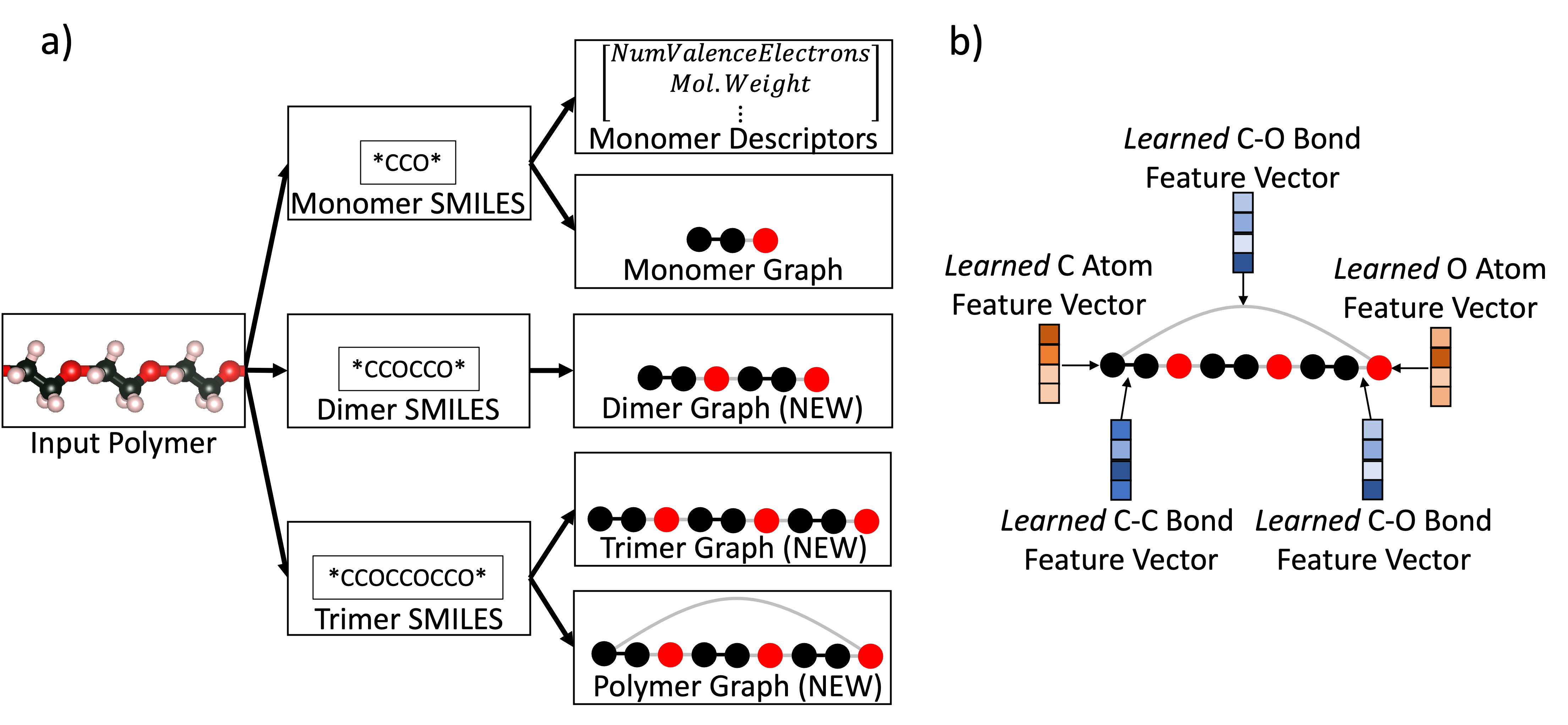}
\includegraphics[width=\textwidth]{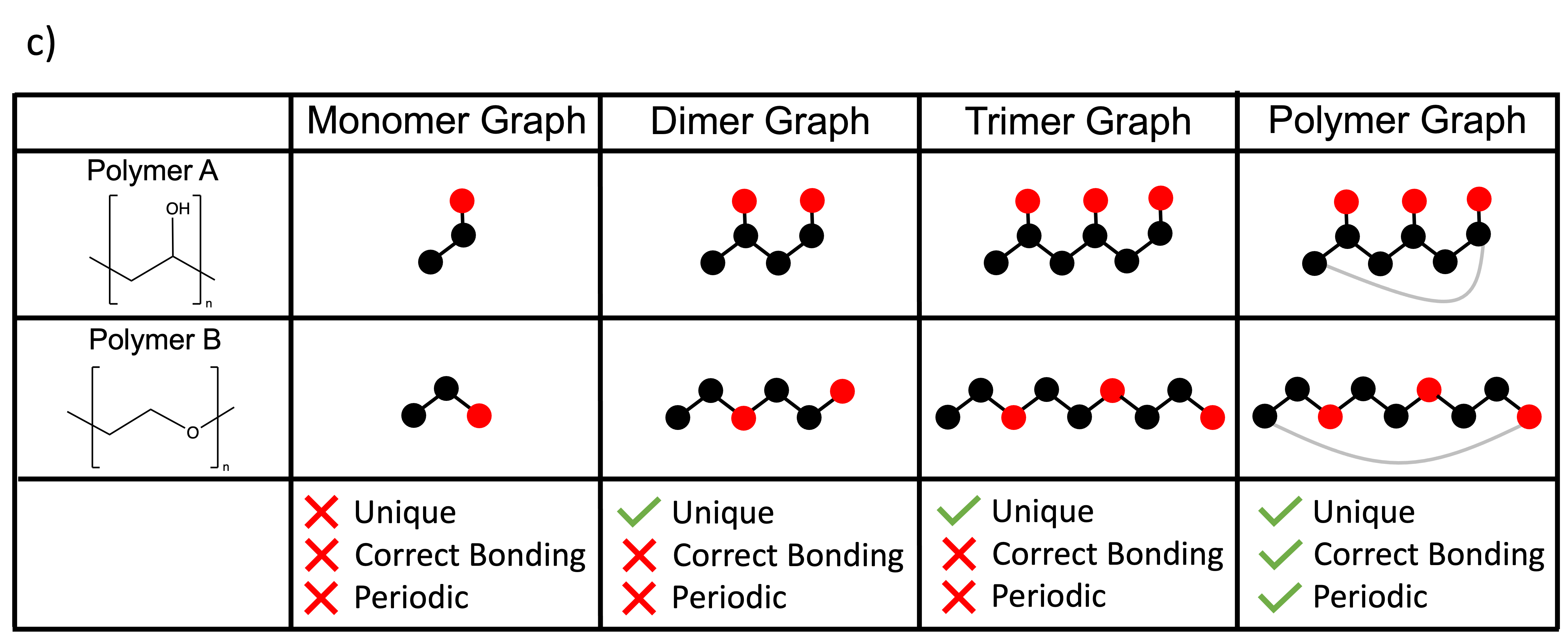}
\caption{ a) 5 polymer representations considered in this work. Monomer graphs and chemical descriptors of the monomer have been reported previously, whereas dimer graphs, trimer graphs, and \textit{polymer graphs} are novel representations developed in this work. b) Structure of the \textit{polymer graph} representation. All atoms (nodes) and bonds (edges) in the graph are represented by learned feature vectors. In this example, the additional periodic edge is featurized in the same manner as all other C-O edge vectors.  c) Representative graphs for two distinct polymers. Carbon and oxygen atoms are depicted by black and red circles, respectively. For clarity, hydrogen atoms are omitted. Generating monomer graphs does not produce a unique representation and results in incorrect functional groups. Dimer and trimer graph representations are unique for Polymers A and B, but result in incorrect bonding environments on the atoms at the end of the chain. The \textit{polymer graph} representation developed in this work creates distinct representations for Polymers A and B, while ensuring the correct bonding environment for terminal atoms of the polymer chain and capturing the periodicity of the polymer.}
\label{fig:figure_1_v1}
\end{figure}

Using monomer-based representations as inputs for ML algorithms results in at least three notable issues: i) the generation of non-unique representations for polymers that are composed of the similar monomer repeat units (see "monomer graphs" of Polymer A and B in Figure 1 c), ii) incorrect bonding environment of the polymer, and iii) a lack of periodicity. Polymers A and B in Figure 1c) are two distinct polymers with notably different chemical structures and properties. Since the monomer of a polymer is defined as the smallest repeat unit of the polymer chain, the monomer of the two polymers in Figure 1 c) is therefore the same C-C-O subunit (or CH$_3$CH$_2$OH if including hydrogen atoms). Regardless of the specific details of the featurization used, generating a representation of Polymer A and B based only on their monomers will result in two identical representations (see Supplementary Note 1). In turn, an ML algorithm that then uses these representations to learn a structure-property relationship will, at best, only be able to learn the average value of the property across the polymers, which may lead to large prediction errors. The second issue that will arise from the use of monomer-based representations is the incorrect featurization of the polymer. Polymer B consists of a repeating (-C-C-O-) backbone, which is characterized by an ether functional group (R-O-R, where R is any alkane). However, the monomer of Polymer B ( C-C-O) does not contain an ether functional group since this ether group is artificially fragmented when forming the monomer representation. This is particularly problematic since both fingerprinting approaches and property descriptors utilize the presence of functional groups and connectivity in their representations (Supplementary Note 1).\cite{landrum_g_rdkit_2006,morgan_generation_1965,rogers_extended-connectivity_2010} Thirdly, the failure to include periodicity in these monomer-based representations results in representations that are inconsistent due to trivial translations along the polymer backbone. The polymer (-C-C-O-)$_n$ can also be written as (-C-O-C-)$_n$ by translating all atoms one position along the polymer chain. However, the monomer representations for these two equivalent polymer representations would be C-C-O and C-O-C, respectively. As shown in this example, the failure to explicitly account for periodicity in these monomer-based representations can result in inconsistent featurizations, which can negatively impact the ability of ML algorithms to learn the underlying structure-property relationships. These inconsistent featurizations are illustrated in Supplementary Note 1.
\\

Fundamentally, all three of the aforementioned problems arise due to a failure to account for the periodic nature of polymers. In other words, a finite molecule (i.e. a monomer) cannot be used to represent a continuous, extended chain of atoms without fragmenting chemical bonds in an unphysical manner. We note that featurizing Polymers A and B by generating dimers (or any larger oligomer chain) instead of monomers results in unique representations. Using a dimer representation, however, still results in incorrect featurizations due to the incomplete bonding environment of the atoms at the terminal end of the dimer. In this work, we introduce a new polymer representation that we refer to as a \textit{polymer graph} that solves all three of these problems. The core idea of the \textit{polymer graph} is that adding a bond between the two terminal ends of the polymer chain creates a periodic polymer representation (Figure 1b)). The periodic nature of this representation eliminates the need for artificially cleaving chemical bonds, ensuring that the correct bonding environment is maintained for all atoms in the polymer. We note that this approach is strongly reminiscent of the \textit{crystal graph} representations that has had wide success in representing periodic crystalline materials that extend indefinitely in all three dimensions.\cite{xie_crystal_2018} \textit{Crystal graph} representations have solved the problem of representing an infinite 3-D crystal by treating it as the smallest repeat unit (referred to as the unit cell) that then transmits chemical information by interacting with periodic copies of itself in all directions. In an analogous fashion, the \textit{polymer graph} developed in this work is formed from the repeat unit of the polymer that is able to transmit chemical information through the added periodic bond. We emphasize that this periodic bond is used only for representing the polymer, but should not be interpreted to mean that the polymers are always part of a chemical ring system.
\\

Graph-based representations of molecules and crystalline solids have recently been the subject of considerable research efforts and are currently the state-of-the-art representation for predicting the properties of these systems.\cite{gasteiger_gemnet_2022,dimenet_2022,chemprop_yang_analyzing_2019} Generally these graphs represent each atom in the chemical system, \textit{$i$}, by an atom embedding vector (\textit{h$_i$}) which is updated through messages that are sent through the edges of the graph. These update messages typically operate on all atoms and bonds in the local neighborhood of atom \textit{$i$}, where the local neighborhood is given by the edges of the graph that are connected to atom \textit{$i$}. To generate our \textit{polymer graph} representation, we first create a standard trimer graph and then add an additional edge to connect the two terminal atoms of the trimer (Figure 1 b)). This added edge in our \textit{polymer graph} representation not only maintains the correct bonding environment for all atoms, but also serves to ensure that messages (i.e. chemical information) can be sent throughout the polymer graph. For example, the left-most carbon atom in the monomer graph in Figure 1a) will only directly receive messages from the central carbon atom since this is the only atom that it is bonded to. This is not an accurate representation of the true chemical structure of this polymer since this carbon atom is directly bonded to both a carbon atom and an oxygen atom. On the other hand, since our \textit{polymer graph} representation creates an edge between the terminal carbon and oxygen atoms, messages are directly passed between both of these two atoms, thus allowing the atom embedding vector of the terminal carbon atom to more accurately represent its true chemical environment. We construct the \textit{polymer graph} from a trimer graph (rather than a dimer or monomer graph) to ensure that the added periodic bond is between atoms that are distinct and not already bonded to each other.
\\

\subsection*{Performance of Graph Representations}

To demonstrate the accuracy and generality of our \textit{polymer graph} representation for predicting polymer properties, we collect a dataset of polymer properties containing 15,219 datapoints across 10 distinct properties (see Methods for dataset details). Notably, this dataset includes a diverse set of property types (thermal, structural, optical, electronic) as well as data collected from both experimental and computational sources, which allows us to assess the capability of our approach to learn meaningful polymer representations for general property prediction tasks. To perform a baseline comparison, we also train ML models on all 10 properties with two commonly used monomer-based representations: i)  molecular descriptors and ii) molecular graphs of the monomer (Table 1). All graph-based models are trained with a MPNN model, whereas a random forest (RF) model is used for training with the monomer descriptors (see Methods for model architecture and hyperparameter details).
\\

\setlength{\heavyrulewidth}{1.5pt}
\setlength{\abovetopsep}{4pt}
\begin{table}[!htbp]
\centering
\caption{Performance of random forest (RF) and message-passing neural network (MPNN) models for predicting polymer properties. For each property, the reported error is the root mean square error (RMSE) on the validation set of 5-fold cross validation. The mean normalized RMSE is obtained by normalizing the RMSE of a model against the RMSE of the RF model and then averaging across all 10 properties. The best performing model for each property is shown in bold.}
\begin{tabular}{*6c}
\toprule
\thead{Property} & \thead{Unit} &  \thead{Number of \\Data Points} & \thead{RF/\\ Molecular Descriptors} & \thead{MPNN/\\Monomer Graph} & \thead{MPNN/\\Polymer Graph} \\
\toprule
Atomization Energy & eV/atom & 390 & 0.100 & 0.057 & \textbf{0.038} \\
\hline
Crystallization Tendency & \% & 432& \textbf{18.050} & 19.850 & 20.180 \\
\hline
Band Gap Chain & eV & 3380 & 0.579 & 0.528 & \textbf{0.449} \\
\hline
Band Gap Bulk & eV & 561 & 0.698 & 0.581 & \textbf{0.528} \\
\hline
Electron Affinity & eV & 368 & 0.450 & 0.306 & \textbf{0.289}\\
\hline 
Ionization Energy & eV & 370 & 0.456 & 0.473 & \textbf{0.385}\\
\hline
Refractive Index & - & 382 & 0.104 & 0.091 & \textbf{0.090}\\
\hline
Dielectric Constant & - & 382 & 0.593 & \textbf{0.533} & 0.535\\
\hline
Glass Transition Temperature & K & 7235 & 41.39 & 38.79 & \textbf{34.91}\\
\hline
Density & g/cc & 1719 & 0.099 & 0.090 & \textbf{0.086}\\
\bottomrule
Mean Normalized RMSE & - & - & 1.000 & 0.875 & \textbf{0.800}\\
\bottomrule
\end{tabular}
\end{table}

Comparing the performance of the RF/Molecular Descriptors and the MPNN/Monomer Graph in Table 1 illustrates the increase in performance that is realized when using graph-based models for polymer property prediction. Across 8 out of 10 properties, the MPNN/Monomer graph model outperforms the RF/Molecular Descriptors model. On average, the MPNN/Monomer Graph model achieves 12 \% lower error than the RF/Molecular Descriptors model, with remarkably large reductions in error for predicting atomization energy (43\%), band gap bulk (17\%), and electron affinity (32\%). Both models perform similarly at predicting ionization energy, whereas the RF/Molecular Descriptors model performs slightly better only at predicting the crystallization tendency. The poor performance of the monomer graph model for predicting crystallization tendency may be due to the fact that these values were calculated with the use of an approximate group contribution method.\cite{venkatram_predicting_2020} The low fidelity of this dataset may therefore be preventing the graph model from learning reasonable atom/bond features for predicting crystallization tendency. Based on the results across all properties, we conclude that graph-based models can generally be expected to perform, at worst, comparably to RF/Molecular Descriptor models, with the potential to perform significantly better on certain properties. This finding is in agreement with a previous report which showed that monomer graph-based models outperform ECFP molecular descriptors for predicting the glass transition temperature, melting temperature, density, and elastic modulus for a subclass of polymers known as polyamides. \cite{park_graph_prediction_2022} In the present work, we expand this finding to show that graph-based models can generally be expected to outperform models based on molecular descriptors across a broader range of polymer properties and for datasets that contain all subclasses of polymers. 
\\

\subsection*{Role of Graph Construction on Predictions}
Now that we have demonstrated the ability of graph-based models to perform well at predicting polymer properties, we now explore how the construction of the graph itself can influence model performance. Table 1 illustrates the performance of monomer graph and \textit{polymer graph} representations with the same MPNN model. Although we have already demonstrated that the MPNN/Monomer Graph model considerably outperforms the current state-of-the-art RF/Molecular Descriptor models, we find that using our \textit{polymer graph} representation results in a further 10\% average reduction in error compared to the monomer graph model and does not perform significantly worse than the monomer graph model for any of the properties. The \textit{polymer graph} representation results in a significant reduction in error for predicting atomization error (33\%), band gap chain (15\%), ionization energy (19\%) and glass transition temperature (10\%), compared to the monomer graph representation. Since the improvement in property prediction accuracy achieved by the \textit{polymer graph} is highly property-dependant, we note that the 10\% average reduction in error reported here is likely to vary depending on the chosen set of properties of interest. Nevertheless, we do not find any properties where the \textit{polymer graph} performs significantly worse than the monomer graph representation, leading us to conclude that the \textit{polymer graph} should be expected to yield more accurate property predictions across a wide range of properties. 
\\

Our original hypothesis to motivate the development of our \textit{polymer graph} representation was that it can improve over monomer graph representations by being able to distinguish between polymers that have the same monomer representation. To test this idea, we construct a test set of 179 glass transition temperature (\textit{T$_g$}) datapoints where each polymer in the dataset has at least one other polymer that contains the same monomer representation, as depicted by polymers A-C in Figure 2. We train a MPNN/Monomer Graph model and a MPNN/Polymer Graph model on a \textit{T$_g$} training set that does not include these 179 polymers and then evaluate their ability to predict the \textit{T$_g$} of the 179 polymers in the test set. This experiment therefore allows us to quantify the impact on prediction accuracy that results from the inability of monomer graphs to distinguish between polymers that have the same monomer representation.  
\\

\begin{figure}
\centering
\includegraphics[width=\linewidth]{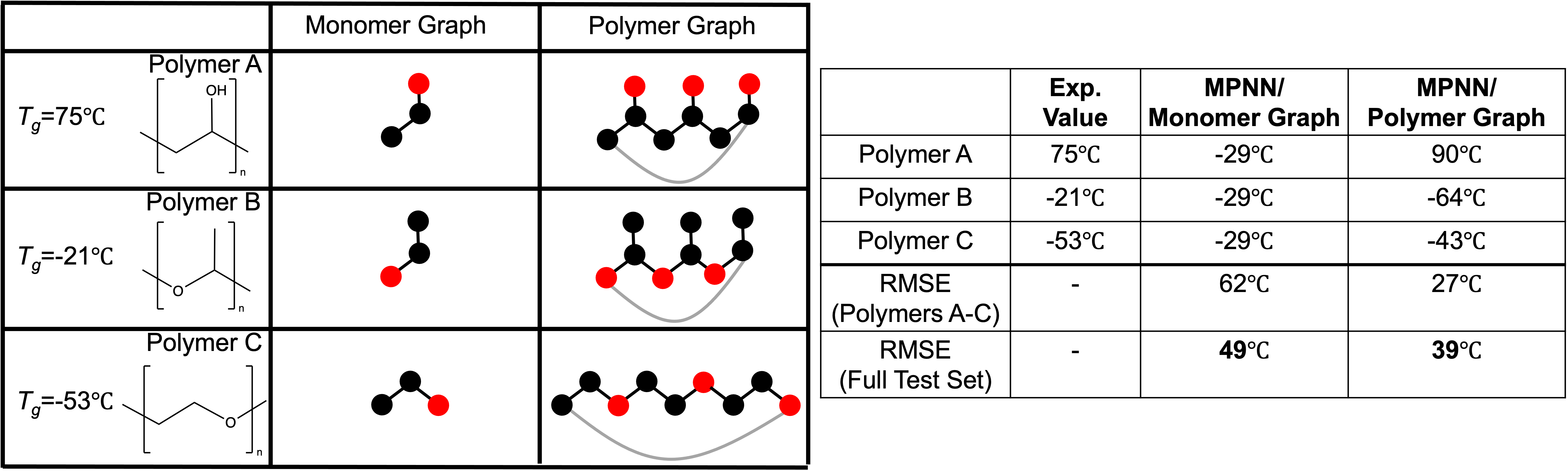}
\caption{Comparison of the performance of monomer graph and \textit{polymer graph} representations for predicting \textit{T$_g$} of a test set of polymers that contain at least one other polymer with the same monomer structure.}
\label{fig:monomer_test_set_figure_v1.png}
\end{figure}

As illustrated in Figure 2, the inability of the MPNN/Monomer graph model to distinguish between polymers A-C results in the same \textit{T$_g$} prediction for all three polymers. Since the experimental \textit{T$_g$} value varies largely between these three polymers, this results in an unacceptably large prediction error of over 60$^{\circ}$C. Although polymers A-C are chosen as an illustrative example, a more statistically significant comparison can be obtained from the performance on the entire test set of 179 polymers. Consistent with our original hypothesis, we see that the inability of the monomer graph representation to distinguish between the polymers in this test set results in significantly worse performance when compared to Table 1 (49$^{\circ}$C vs. 39$^{\circ}$C). On the other hand, the \textit{polymer graph} representation is not only able to distinguish between these polymers, but also quantitatively capture the trends in \textit{T$_g$} values for polymers A-C with a RMSE that is nearly 60\% less than that of the monomer graph. These results therefore provide a partial explanation for the superior of performance of the \textit{polymer graph} representation in Table 1, relative to the monomer graph representation. By being able to distinguish between polymers of the same monomer structure, the \textit{polymer graph} representation is therefore better able to learn quantitative structure-property relationships for the properties in Table 1. Since the proportion of polymers that have the same monomer structure is relatively small (~2\% of the \textit{T$_g$} dataset), we anticipate that this effect is currently small, but will grow in importance as more polymers are discovered. 
\\

The results of the experiment shown in Figure 2 also have important practical implications for the reliability that can be expected when using ML methods to discover new polymers. The \textit{T$_g$} of a polymer is the temperature at which the thermal motions of the polymers are large enough to overcome the forces that hold the amorphous polymer together, such that the polymer is able to undergo large-scale molecular motions.\cite{bicerano_prediction_2002} Accurately predicting the glass transition temperature is therefore of great importance since \textit{T$_g$} is often treated as the maximum temperature that the polymer can be used.\cite{hale_glass_1991} Despite having identical chemical compositions, the three polymers shown in Figure 2 exhibit a remarkably broad range of experimental \textit{T$_g$} values due to their unique chemical structures. In particular, the presence of the hydroxyl (OH) group on Polymer A leads to an increase in interchain attractive forces, resulting in a significantly larger  \textit{T$_g$} value than for Polymers B and C.\cite{feldstein_relation_2001} The inability of monomer-based graph representations to distinguish between these polymers therefore severely hurts the trustworthiness of the model. If one were to use the MPNN/Monomer Graph model to discover new polymers, the large errors in this subset of \textit{T$_g$} predictions could manifest in the discovery of polymers that are thermodynamically unstable at the desired operating temperature. By comparison, the consistent performance seen by the MPNN/Polymer Graph model helps to instill confidence that the predictions will be robust across a large dataset of polymer materials.
\\

\subsection*{Ablation Studies on Periodic Graphs}
Although the results in Table 1 and Figure 2 suggest that the \textit{polymer graph} representation can more accurately predict polymer properties than both the monomer graph and molecular descriptor representations, they do not fully explain what is causing this improvement in performance. Understanding specifically how polymer representation can impact predictive performance is important for the development of increasingly better polymer representations. To accomplish this goal, we perform several ablation experiments whereby we systematically alter components of the \textit{polymer graph} representation to allow us to quantify the contribution of each component to overall model performance (Figure 3). We conduct this experiment by training the same MPNN model as before on the same 10 properties listed in Table 1. For clarity, we illustrate the performance of each representation by calculating the average performance across all properties (Figure 3). The property-specific performance of each representation is given in Table S1.
\\

\begin{figure}
\centering
\includegraphics[width=\linewidth]{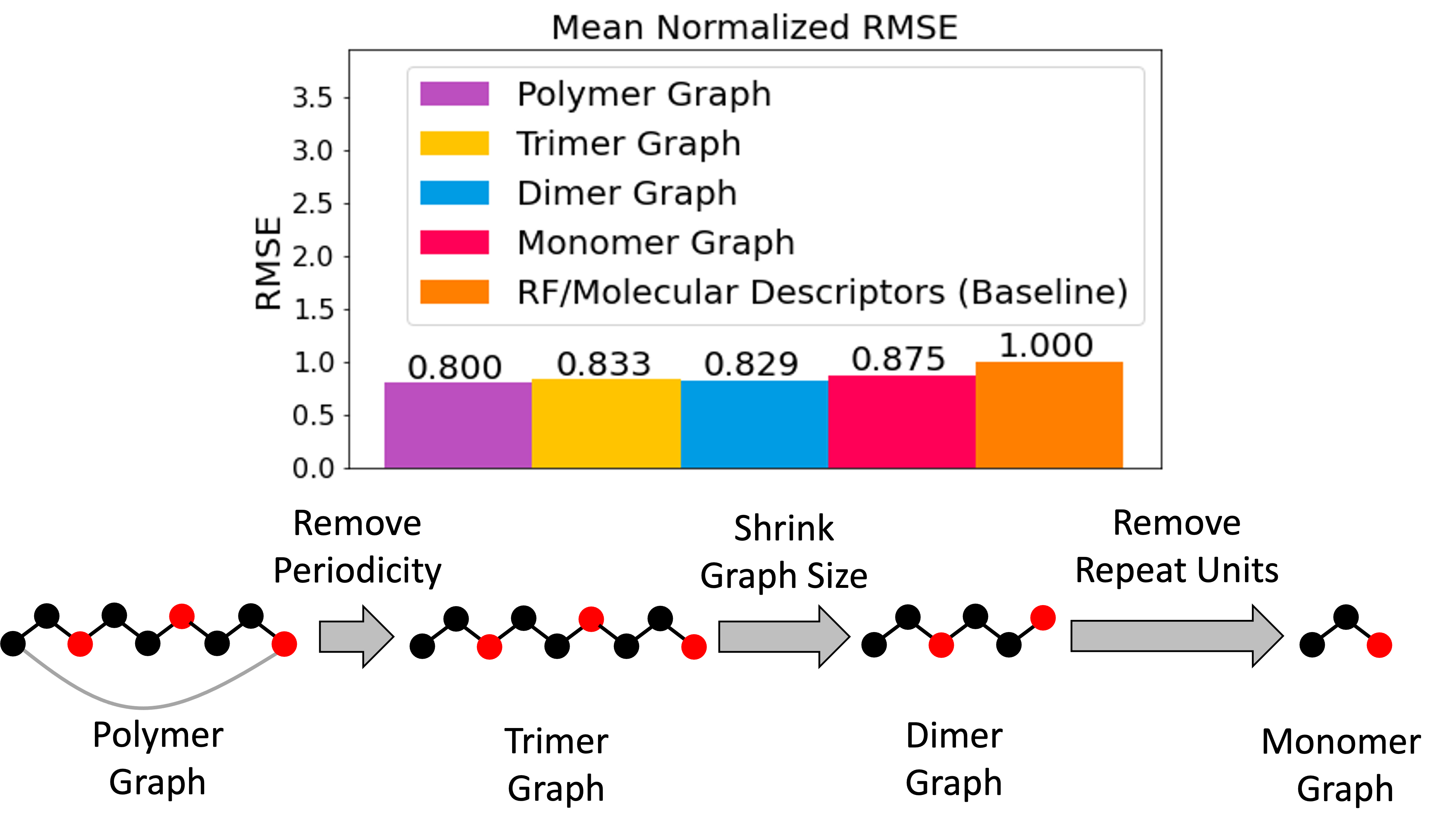}
\caption{Results of ablation experiments for various graph-based polymer representations. The mean normalized RMSE is calculated for all 10 properties listed in Table 1 and normalized against the performance of the RF/Molecular Descriptors representation. All property-specific RMSE values are given in Table S1.}
\label{fig:figure_2_v3}
\end{figure}

The first ablation experiment that we perform is to remove the periodic edge in the \textit{polymer graph} representation, which results in a trimer graph. This ablation experiment highlights the importance of including periodicity in polymer representations. Although the \textit{polymer graph} and trimer graph representations only differ by a single edge in their graphs, the inclusion of  periodicity in the \textit{polymer graph} representation results in a 4\% average reduction in performance error. Most notably, a 32\% reduction in the error of atomization energy predictions is observed when using the \textit{polymer graph} representation instead of the trimer graph (Table S1).  Next, we explore the effect of increasing the size of the polymer chain that is used in the graphs by comparing the performance achieved when using the trimer graph and dimer graph representations. We do not observe any significant performance difference between these two representations. This result is perhaps unsurprising since the trimer graph does not provide any additional chemical information compared to the dimer graph representation. Based on this finding, we anticipate that using even larger graph representations (i.e. tetramer or pentamer) will not improve model performance, but will unnecessarily require more computational resources for model training. Finally, comparing the performance of the dimer graph and the monomer graph representations illustrates that including more than one repeat unit in the graph significantly improves performance. It may seem counter-intuitive that the trimer graph and dimer graph representations perform similarly, whereas dimer graph representations considerably outperform monomer graph representations. However, this result can be explained by noting that the dimer graph representation contains additional chemical bonding information, relative to the monomer graph, since it directly represents the atoms that are bonded together to form the polymer chain. On the other hand, the trimer/dimer graph representations do not differ in the chemical information that they contain, but are simply scaled versions of each other.      
\\

\subsection*{Analysis of MPNN Learned Polymer Descriptors, Hand-Designed Descriptors and Chemical Intuition}
We note from the ablation experiments illustrated in Figure 3 that the largest reduction in error relative to the RF/Monomer Descriptors baseline is due to using graph-based representations with the MPNN model. Using graph-based models results in at least a 13\% average reduction in error, relative to the RF/Monomer Descriptors baseline, whereas further refinements to the graph representation only affects the average performance by approximately 3-5\%.  To better understand why using graph-based representations in combination with a MPNN architecture leads to such a significant performance improvement, we analyze the learned feature vectors that are generated from the MPNN model. The MPNN model utilized in this work uses learned bond and atom features that are aggregated into a single fixed-length vector. This aggregated vector is then passed through a feed-forward neural network with two hidden layers to generate the final property prediction. As a result, the vector in the final hidden layer of this feed-forward neural network can be considered as a learned encoding of the polymer that has been optimized by the MPNN model to make accurate property predictions. These vector encodings can therefore be analyzed to determine what chemical information the MPNN model is using to generate its predictions.
\\

Whereas the RF/Monomer Descriptors baseline used previously (Table 1) utilizes pre-determined molecular descriptors (such as molecular weight), the vector encodings extracted from our MPNN model are vector representations that have been specifically optimized for the particular property that they are trained on. Comparing the performance of the RF/Monomer Descriptors baseline and a RF model trained with these learned encodings therefore allows us to quantitatively understand if the superior performance of the graph-based models are due to these optimized encodings or the MPNN model architecture. To perform this experiment, we train the MPNN model on \textit{k-1} folds (\textit{k=5}) of the \textit{T$_g$} dataset. This trained MPNN model is then used to generate vector encodings for the polymers in the same \textit{k-1} folds of the dataset. Then, these vector encodings are used as inputs to train the RF model on the same \textit{k-1} folds of the dataset that the MPNN model was trained on. Finally, the test performance of the RF model is calculated for the remaining fold of the dataset that was not seen by the MPNN model nor the RF model in training. This procedure is repeated for \textit{k} iterations to ensure that each fold of the dataset is used as the test set once and the reported performance is averaged across these \textit{k} iterations. This training procedure prevents data leakage by ensuring that the test set is not seen by the MPNN model or the RF model before determining test performance.  
\\

The results of this experiment are given in Figure 4a). Remarkably, we find that the RF model with learned descriptors performs comparably to the MPNN/Polymer Graph model at predicting \textit{T$_g$}. This result suggests that the excellent performance of the MPNN/Polymer Graph model is not due to MPNNs being a more powerful model architecture. Rather, the learned descriptors that the MPNN model extracts provides a more meaningful representation of polymers, leading to superior predictive capability. This finding is exciting because it suggests that the use of learned polymer representations that better capture the underlying chemical properties of polymers can lead to greatly improved predictive capabilities, even with relatively simple model architectures like random forests. It is important to note that the predetermined molecular descriptors that have been previously used for polymer prediction tasks were first developed as general descriptors for molecules, but with no guarantee that these descriptors will contain properties that are relevant for describing polymer properties.\cite{landrum_g_rdkit_2006} On the other hand, these learned MPNN descriptors are specifically trained to capture the relevant components of a polymer's structure that give rise to its properties. In this way, the MPNN model explored in this work provides a recipe for automatically extracting meaningful polymer representations directly from polymer property datasets in a manner that does not require any hand-crafted featurization.
\\

Although we have shown that the MPNN model is capable of extracting features that achieve better performance than predetermined features, it is nonetheless important to gain an understanding of what the MPNN learned features physically represent. To gain this chemical insight, we calculate the correlation between the 5 heaviest weighted features of the final layer of the MPNN/Polymer Graph model and the predetermined molecular descriptors for the polymers in the \textit{T$_g$} dataset. The 5 molecular descriptors that are most correlated with the MPNN learned features for the MPNN/Polymer Graph model are depicted in Figure 4b). Two of the most important known structural features for predicting \textit{T$_g$} in polymers is: i) the presence of polar functional groups such as NH$_2$ and OH that increase intermolecular forces, resulting in a higher \textit{T$_g$} value and ii) the stiffness of the polymer chain, which is typically described in terms of the presence of rigid ring structures in the polymer backbone.\cite{bicerano_prediction_2002} The molecular descriptors shown in Figure 4b) are colored according to which one of these two known structural factors they explicitly capture. Interestingly, we see that the majority of the learned MPNN features are strongly correlated with these known relevant chemical features, which suggests that our MPNN model is autonomously learning the chemical features that are responsible for \textit{T$_g$} directly from the dataset of \textit{T$_g$} values. In particular, the most heavily weighted MPNN feature is correlated with the presence of polar groups, whereas the other top 5 features provide complex descriptions of the presence of ring structures in the polymer. However, we emphasize that these MPNN polymer encodings also contain new information that is not captured by the predetermined descriptors, as is shown by the superior performance of the RF model when using learned descriptors instead of predetermined descriptors (Figure 4a).  
\\

\begin{figure}
\centering
\includegraphics[width=\linewidth]{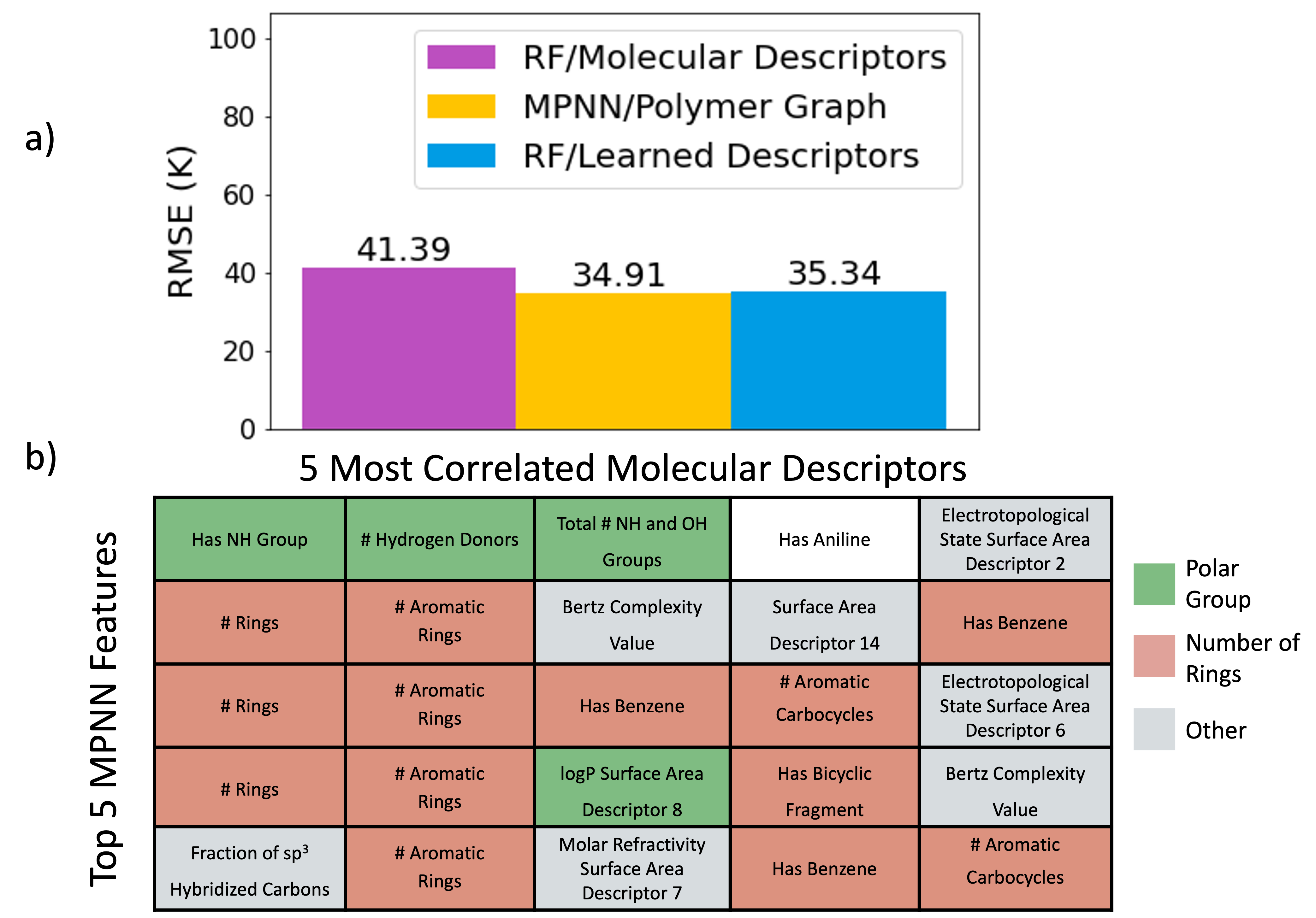}
\caption{ a) Comparison of RF models with predetermined molecular descriptors and learned descriptors against the MPNN/Polymer Graph model for predicting glass transition temperature. b) Correlation between the 5 most heavily weighted features in the last layer of the MPNN model and the predetermined molecular descriptors. Features are colored according to the type of chemical information that they capture. For clarity, the molecular descriptors from Rdkit have been renamed to more explicitly describe their property. The original Rdkit molecular descriptors names and an explanation of these descriptors are given in Supplementary Note 2.}
\label{fig:figure_3_v5}
\end{figure}

\section*{Discussion}

Generally, this work has demonstrated how the development of representations that better account for the unique and complex chemical structure of polymers can be a powerful strategy for increasing the accuracy of polymer property predictions. Through a number of illustrative examples, we have demonstrated how a failure to account for periodicity in polymers can lead to property predictions that are inaccurate and unreliable. We have shown that these issues can be mitigated by our novel \textit{polymer graph} representation that explicitly accounts for the inherent periodicity of polymeric materials. The results presented here therefore demonstrate that utilizing chemical intuition to construct physically-informed polymer representations is a viable strategy for advancing the predictive capabilities of ML models.  

The polymer informatics community has benefited greatly from the adaptation of machine learning algorithms and representations that were first developed for molecules. This approach is based on the assumption that general molecular descriptors are likely to work well across a broad spectrum of chemical classes, including polymers. In this work, we test this assumption by exploring the improvement in property predictions that can be achieved through the use of learned descriptors that are specifically optimized to capture the unique structural characteristics of polymers. We have shown that utilizing polymer-specific descriptors results in considerable improvements in the accuracy of polymer property predictions, compared to standard molecular descriptors. These results suggest that the structural characteristics that give rise to polymer properties are distinct from general molecular descriptors. This is perhaps unsurprising since it is well known that polymers are affected by structural elements that do not have molecular analogues, such as the branching structure of the polymer or its interchain interactions. We anticipate that the continued development of representations that are customized to capture the unique and complex structure of polymeric systems is therefore a fruitful direction for further pushing the limits of predicting polymer properties with ML.

\section*{Methods} \label{Methods}

\subsection*{Data}
The polymer dataset used in this work consists of 2 experimentally measured properties that were manually obtained from the PolyInfo database (glass transition temperature (\textit{T$_g$}), and density.\cite{otsuka_polyinfo_2011} The data for the other 8 properties (atomization energy, crystallization tendency, band gap chain, band gap bulk, electron affinity, ionization energy, refractive index, dielectric constant) were previously calculated with density functional theory (DFT).\cite{kuenneth_polymer_2021} Altogether, this dataset consists of 15,219 total datapoints, spanning 9,935 unique homopolymer structures. For the experimentally measured properties, we exclude all measurements that correspond to compound or composite polymer samples. After these measurements are excluded, the experimental property value for each polymer is taken to be the median of all measurements.

\subsection*{Molecular Descriptors}
The molecular descriptors used in combination with the random forest model were obtained from the open source cheminformatics tookit, RDKit.\cite{landrum_g_rdkit_2006} In total, these descriptors contain 200 unique features: 115 of which describe chemical characteristics (such as the number of rings or molecular weight), while the other 85 are binary features that describe the presence of molecular fragments (such as the presence of an amide functional group). For each property, any features that have a constant value across the whole dataset are removed prior to training. The molecular descriptors are generated from the monomer SMILES string of the polymer that has the '*' characters removed. For example, the monomer input SMILES string for the polymer (-CH$_2$-CH$_2$-O-)$_n$ would be 'CCO'.

\subsection*{Models}
The Message Passing Neural Network (MPNN) used in this work is implemented in the Chemprop package.\cite{chemprop_yang_analyzing_2019} Notably, this network utilizes directed bond-level messages (rather than the atom-level messages used in traditional MPNNs) to transmit chemical information across the molecule.  This network architecture operates on input molecular graphs to first generate a vector encoding of the graph, followed by a feed-forward neural network that converts the graph encodings into property predictions. The size of the bond message vectors is set to 300 and a total of 3 message-passing steps are used. The feed-forward neural network consists of 2 hidden layers, each consisting of 300 hidden neurons. All hyperparameters used are the Chemprop default values. We point interested readers to the Chemprop documentation for an extensive list of all hyperparameters used in this architecture.\cite{chemprop_yang_analyzing_2019}  Unless otherwise stated, all models are trained with 5-fold cross-validation (CV) for 50 epochs on each fold. The best-performing model is chosen by selecting the model with the lowest root-mean-square error (RMSE) performance on the validation set. The Random Forest (RF) model was implemented in scikit-learn version 0.24.1 with 500 trees.\cite{scikit-learn}

\bibliography{sample}

\section*{Acknowledgements}

This work was performed under the auspices of the U.S. Department of Energy by Lawrence Livermore National Laboratory under Contract DE-AC52-07NA27344.






\newpage

\section*{Supplementary Information}
\subsection*{Supplementary Note 1: Featurization of Monomers}

\begin{figure}[!htb]
\centering
\caption*{\textbf{Figure S1.} Visualization of molecular descriptors generated in RDKit for 4 representative SMILES strings for polyethylene glycol, (-CH$_2$-CH$_2$-O-)$_n$. The molecular property features are normalized by the maximum value that they have in the full dataset of \textit{T$_g$} values.}
\includegraphics[width=\linewidth]{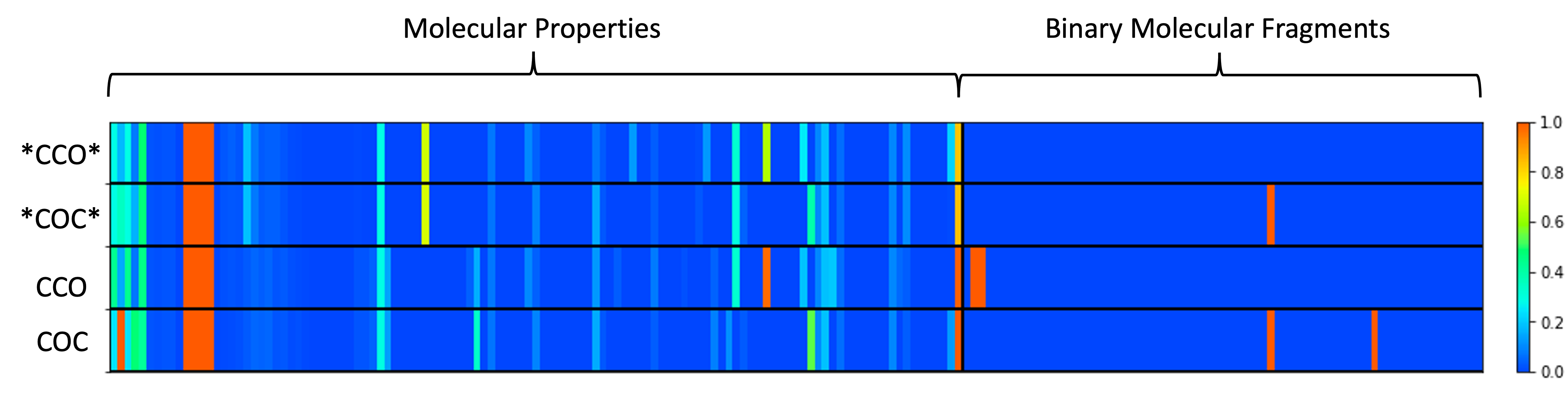}
\label{fig:si_featurization_errors_figure_v1}
\end{figure}

To illustrate the inconsistencies that arise when using preexisting molecular descriptors to represent polymers, Figure S1 provides a visualize of the molecular descriptors generated from RDKit for 4 representative SMILES strings. According to the syntax of the SMILES strings, the '*' character is used to indicate the connection between repeating units of a polymer. The SMILES notation also does not explicitly indicate hydrogen atoms, but instead assumes that all atoms have the number of hydrogen atoms required for full valency. The '*CCO*' SMILES string therefore represents a (-CH$_2$-CH$_2$-O-)$_n$ polymer and the '*COC*' SMILES string represents the same polymer, (-CH$_2$-O-CH$_2$-)$_n$, that has been trivially translated along the polymer backbone. Even though '*CCO*' and '*COC*' represent the same polymer, Figure S1 illustrates the different feature vectors that are generated from these SMILES strings. An understanding of why this difference arises can be obtained by looking at the molecular fragment features that are generated for these two SMILES string. The only difference in the molecular fragment features is the 'fr\textunderscore ether' feature, which is a binary feature that indicates the presence of an ether functional group. As shown in Figure S1, the '*CCO*' has fr\textunderscore ether=0 (indicating the absence of an ether), whereas '*COC*' has fr\textunderscore ether=1 (indicating the presence of an ether). Evidently, the inconsistent featurization between these two chemically identical SMILES strings is due to a lack of periodicity. Since the '*CCO*' SMILES string does not treat the terminal oxygen atom as being bonded to two neighboring carbon atoms, using the '*CCO*' SMILES string results in an incorrect featurization of fr\textunderscore ether=0, when this polymer does in fact contain an ether.

The impact of including the '*' character in the SMILES representation is illustrated with feature vectors of the 'CCO' and 'COC' SMILES strings, which are the monomer subunits of '*CCO*' and '*COC*', respectively. Interestingly, all four of these inputs yield four different feature vectors. Although these inconsistencies could be accounted for by designating a standardized SMILES string to use as an input for a given polymer, it is unlikely that such a standardized system would always capture the correct periodicity of the polymer. Instead, the use of the \textit{polymer graph} representation as described in the main text ensures that the periodicity of the polymer is maintained and always generates a consistent representation.

\subsection*{Supplementary Note 2: Explanation of Features in Figure 4}

\begin{figure}
\centering
\includegraphics[width=\linewidth]{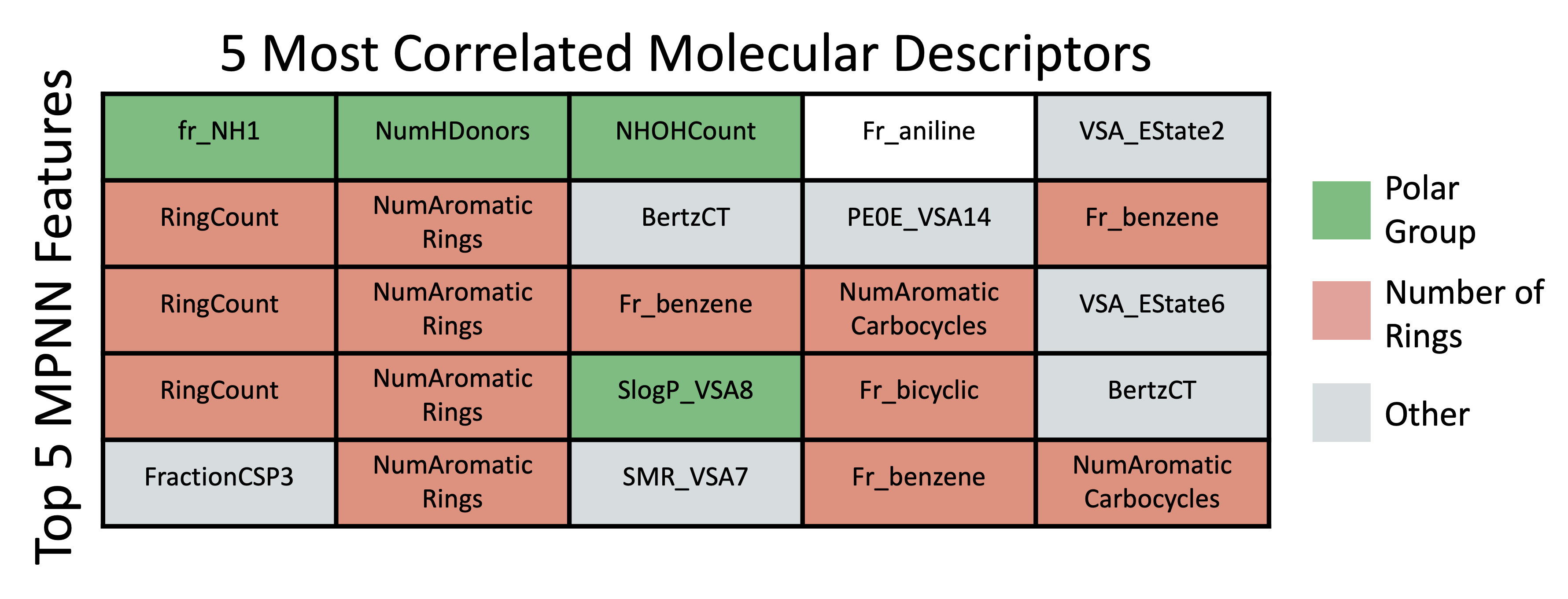}
\caption*{\textbf{Figure S2.} Same as Figure 4b, but molecular features are named according to the Rdkit documentation.}
\label{fig:figure_3_v3_b_only}
\end{figure}

Here, we list the original names of the features (as described in the RDKit documentation). Each feature in the main text is also categorized into capturing 1) the presence of polar groups, 2) the number of rings, and 3) other. For the features that do not explicitly represent these groups, we also provide a simple justification for classifying each feature into each group. 
\\

\textbf{fr\textunderscore NH1}: The presence of a NH1 group in the molecule. Polar Groups.
\\

\textbf{NumHDonors}: Number of hydrogen bond donors in the molecule. Polar Groups.
\\

\textbf{NHOHCount}: The total number of NH or OH functional groups in the molecule. Polar Groups.
\\

\textbf{Fr\textunderscore aniline}: The presence of an aniline functional group in the molecule. Since aniline contains both a polar group (NH) as well as a ring, we classify this feature as capturing both 1) presence of polar groups and 2) the number of rings.
\\

\textbf{VSA\textunderscore EState2}: The electrotopological state of an atom, summed for all atoms in the molecule for a van der Waals surface area between 4.78-5.00. Other.
\\

\textbf{RingCount}: Number of rings in the molecule. Number of rings.
\\

\textbf{NumAromaticRings}: Number of aromatic rings in the molecule. Number of rings.
\\

\textbf{BertzCT}: A topological descriptor that describes the 'complexity' of a molecule. Other.
\\

\textbf{PEOE\textunderscore VSA14}: Approximate descriptor for describing a molecules surface area. Other.
\\

\textbf{Fr\textunderscore benzene}: The presence of a benzene ring in the molecule. Number of rings.
\\

\textbf{NumAromaticCarbocycles}: Number of aromatic carbocyles in the molecule. Number of rings.
\\

\textbf{VSA\textunderscore EState6}: Same as VSA\textunderscore EState2, but calculated for a van der Waals surface area between 6.00-6.07. Other.
\\

\textbf{SlogP\textunderscore VSA8}: The log P value of a molecule is a partition coefficient that describes its lipophilicity. This descriptor describes the number of atoms in the surface area of molecule that have log P values between 0.25 and 0.30. We include this feature in Group 1 since the lipophilicity of a molecule is heavily impacted by the polarity of a molecule.
\\

\textbf{Fr\textunderscore bicyclic}: Presence of bicyclic substructures in the molecule. Number of rings.
\\

\textbf{FractionCSP3}: Fraction of carbons in the molecule that are SP3 hybridized. Other.
\\

\textbf{SMR\textunderscore VSA7}: Molar refractivity (MR) describes the polarizability of a molecule. Similar to the SlogP\textunderscore VSA8 feature, this descriptor describes the number of atoms in the surface area of a molecule that have estimated molar refractivity values between 3.05-3.63. Other.

\setlength{\abovetopsep}{4pt}
\begin{table}[!htbp]
\centering
\caption*{Table S1: Performance of \textit{polymer graph}, trimer graph, dimer graph and monomer graph representations for all 10 polymer properties. All representations are trained with the MPNN architecture described in the Methods section.} 
\begin{tabular}{*7c}
\toprule
Property & Unit &  Num. Data Points & Polymer Graph & Trimer Graph & Dimer Graph & Monomer Graph \\
\toprule
Atomization Energy & eV/atom & 390 & \textbf{0.038} & 0.056 & 0.055 & 0.057 \\
\hline
Crystallization Tendency & \% & 432& 20.18 & 21.14 & 21.06 & \textbf{19.85} \\
\hline
Band Gap Chain & eV & 3380 & \textbf{0.449} & 0.450 & 0.453 & 0.528 \\
\hline
Band Gap Bulk & eV & 561 & 0.528 & \textbf{0.527} & 0.532 & 0.581 \\
\hline
Electron Affinity & eV & 368 & 0.289 & \textbf{0.275} & 0.282 & 0.306\\
\hline 
Ionization Energy & eV & 370 & \textbf{0.385} & 0.435 & 0.435 & 0.473\\
\hline
Refractive Index & - & 382 & 0.090 & \textbf{0.088} & \textbf{0.088} & 0.091\\
\hline
Dielectric Constant & - & 382 & 0.535 & \textbf{0.520} & 0.529 & 0.533\\
\hline
Glass Transition Temperature & K & 7235 & 34.91 & 37.00 & \textbf{34.79} & 38.79\\
\hline
Density & g/cc & 1719 & \textbf{0.086} & 0.088 & \textbf{0.086} & 0.090\\
\bottomrule
\end{tabular}
\end{table}

\end{document}